# A Comparison between Digital Image Watermarking in Tow Different Color Spaces Using DWT2*

Mehdi, Khalili

National Academy of Science
Yerevan, Armenia
e-mail: khalili.mehdi@yahoo.com

**ABSTRACT**

A novel digital watermarking for ownership verification and image authentication applications using discrete wavelet transform (DWT) is proposed in this paper. Most previous proposed watermarking algorithms embed sequences of random numbers as watermarks. Here binary images are taken as watermark for embedding. In the proposed approach, the host image is converted into the YCbCr color space and then its Y channel decomposed into wavelet coefficients. The selected approximation coefficients are quantized and then their four least significant bits of the quantized coefficients are replaced by the watermark using LSB insertion technique. At last, the watermarked image is synthesized from the changed and unchanged DWT coefficients. The experiments show that the proposed approach provides extra imperceptibility and robustness against wavelet compression compared to the traditional embedding methods in RGB color space. Moreover, the proposed approach has no need of the original image to extract watermarks.

**Keywords**
Digital watermarking, discrete wavelet transform, YCbCr color space, LSB insertion technique, imperceptibility, wavelet compression.

## 1. INTRODUCTION

With the rapid growth of Internet technologies and wide availability of multimedia computing facilities, the enforcement of multimedia copyright protection becomes an important issue [1]. Digital media, e.g., images, audio, and video, are readily manipulated, reproduced, and distributed over information networks. These efficiencies lead to problems regarding copyright protection. As a result, creators and distributors of digital data are hesitant to provide access to their digital intellectual property.

Technical solutions for copyright protection of multimedia data are actively being pursued [2, 3]. Watermarking, which allows for the imperceptibly embedding information in an original multimedia data, has widely emerged for copyright protection and ownership identification.

All watermarking methods must satisfy three important requirements: perceptual invisibility, robustness against various images processing, such as compression and geometric distortions (for example cropping), and finally ability of watermarking detection without ambiguity [4].

In this paper digital image watermarking for ownership verification and image authentication applications is presented, where watermark is a binary image that embeds in Y channel of YCbCr color space.

Current techniques described in the literature for the watermarking of images can be grouped into two classes: spatial domain techniques [5, 6] which embed the data by directly modifying the pixel values of the original image, and transform domain methods [7,8] which embed the data by modulating the transform domain coefficients.

A variety of watermarking techniques has been proposed in recent years. One of the earlier watermarking techniques, which used wavelet transform, was based on the adding pseudo random codes to the large coefficients at the high and middle frequency bands of the discrete wavelet transform [9].

The fundamental advantage of our wavelet-based technique lies in the method used to embed the watermark in four bits of lowest significant bits of Y channel DWT approximation coefficients of YCbCr color space of host image using LSB insertion technique.

With this proposed algorithm, we can obtain relatively high payload (almost 20%). Experimental results show the efficiency of proposed techniques among other techniques of digital images watermarking in wavelet transform domain. This results shows that watermarking in YCbCr color space against compression attacks is more robust than watermarking in RGB color space, and has higher transparency in same payload.

## 2. WAVELET TRANSFORF OF IMAGES

The wavelet transform has been extensively studied in the last two decades. Here, we introduce the necessary concepts of the DWT for the purposes of this paper.

The discrete wavelet transform (DWT) is identical to a hierarchical subband system, where the subbands are logarithmically spaced in frequency [10]. For an input sequence of length n, DWT will generate an output sequence of length n.

The 1-D DWT can be completed by the direct pyramid algorithm developed by Mallat [11].

For a 2-D image, a wavelet $\Psi$ and a scaling function $\Phi$ are chosen such that the scaling function $\Phi_{LL}(x, y)$ of low-low subband in a 2-D wavelet transform can be written as $\Phi_{LL}(x,y) = \Phi(x)\Phi(y)$.

Three other 2-dimensional wavelets can also be obtained by using the wavelet associated function $\Psi(x)$ as follows [9, 12, and 13]:

$$\Psi_{LH}(x, y) = \Phi(x)\Psi(y); \text{horizontal}$$
$$\Psi_{HL}(x, y) = \Psi(x)\Phi(y); \text{vertical}$$
$$\Psi_{HH}(x, y) = \Psi(x)\Psi(y); \text{diagonal}$$

Where H is a highpass filter and L is a lowpass filter.

The basic idea for the DWT of a 2-D image is as follows: An image is firstly decomposed into four parts of high, middle, and low frequency (i.e., LL1, HL1, LH1, HH1) subbands by cascading the image horizontally and vertically with critically subsampled filter banks. The subbands labeled HL1, LH1, and HH1 represent the finest-scaled wavelet coefficients. To obtain the coarser-scaled wavelet coefficients, the subband LL1 is further decomposed and critically subsampled. This process is repeatedly a number of arbitrary times which is determined by the application at hand. Fig 1 shows an original image and its DWT decomposition. In this figure, image is decomposed into three levels with ten subbands. Each level has various band-information such as low-low, low-high, high-low, and high-high frequency bands [9].

Note that in Fig 1, the lowest frequency subband is at the top left, and the highest frequency subband is at the bottom right. Furthermore, from DWT coefficients, the original image can be reconstructed. This reconstruction process is called the inverse DWT (IDWT). If I [m, n] represent an image, the DWT and IDWT for I [m, n] can be similarly defined by implementing the DWT and IDWT on each dimension m and n separately [9].

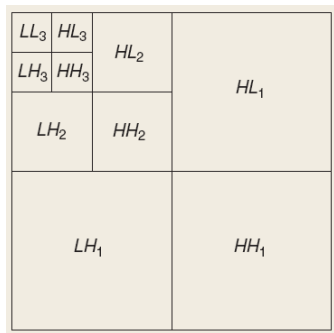

**Fig1. A real case of DWT decomposition.**
**(a) The original image. (b) Its DWT decomposition**

## 3. YCbCr COLOR SPACE

YCbCr color space is used for color images cryptography. In this color space, the Y denotes the luminance component. It means that Y shows the brightness (luma). Also both of Cb and Cr represent the chrominance actors. It means that Cb is blue color minus luma (B_Y) and Cr is red color minus luma (R_Y) [13].

The difference between YCbCr color space and RGB color space is that YCbCr color space represents color as brightness and two color difference signals, while RGB represents color as red, green and blue.

Equations 1 and 2 show transformations between RGB color space and YCbCr color space [14].

$$\begin{bmatrix} Y \\ Cb \\ Cr \end{bmatrix} = \begin{bmatrix} 16 \\ 128 \\ 128 \end{bmatrix} + \begin{bmatrix} 65.481 & 128 & 24.966 \\ -37.0797 & -74.203 & 112 \\ 112 & -93.786 & -18.214 \end{bmatrix} \begin{bmatrix} R \\ G \\ B \end{bmatrix}$$

(1)

$$\begin{bmatrix} R \\ G \\ B \end{bmatrix} = \begin{bmatrix} 0.00456621 & 0 & 0.00625893 \\ 0.00456621 & -0.00153632 & -0.00318811 \\ 0.00456621 & 0.00791071 & 0 \end{bmatrix} \times \left( \begin{bmatrix} Y \\ Cb \\ Cr \end{bmatrix} - \begin{bmatrix} 16 \\ 128 \\ 128 \end{bmatrix} \right)$$

(2)

## 4. PROPOSED WATERMARKING TECHNIQUE

The current study task of digital watermarking is to make watermarks invisible to human eyes as well as robust to various attacks. The proposed watermarking approach can hide visually recognizable patterns in images. The proposed approach is based on the discrete wavelet transform (DWT).

In the proposed approach, the host image is converted into YCbCr channels; the Y channel is then decomposed into wavelet coefficients. Then, we embed watermark in the approximation coefficients of DWT of the host image by modifying four least significant bits. The block diagram of the proposed watermarking approach is shown in Fig 2.

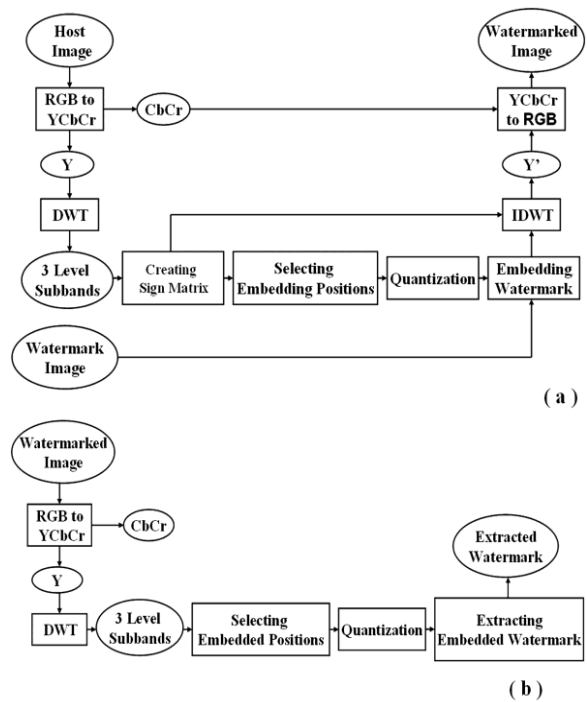

**Fig2. Block diagrams of the proposed watermarking approach. (a) Embedding procedure. (b) Extracting procedure.**

### 4.1. Watermark embedding method

The algorithm for embedding watermark in LL3 coefficients of the host image Y channel is described as follows:

Step 1: Convert RGB channels of a host image W into YCbCr channels using the CCIR 601 standard.
Step 2: Decompose the Y channel into a three-level wavelet pyramid structure with ten DWT subbands, F(H). The coarsest subband LL3 is taken as the target subband for embedding watermarks.
Step 3: Save the signs of selection coefficients in a matrix sign.
Step 4: Quantize absolute values of selection coefficients.
Step 5: Embed watermark W1. For robustness, imperceptibility, and security, the watermark W is embedded in four least significant bits that have smallest quantization error.
Step 6: Effect matrix sign into the embedded coefficients.
Step 7: Reconvert YCbCr channels of a changed host image into RGB channels.
Step 8: A watermarked image W' is then generated by inverse DWT with all changed and unchanged DWT coefficients.
Step 9: Save indexes of changed selection coefficients, and index of the embedded subband as the authenticated key.

## 4.2. Watermark extracting method

The embedded watermark can be extracted using the stored authenticated key after wavelet decomposition of the watermarked image. The extracting process is described as follows:
Step 1: The RGB channels of the watermarked image are converted into YCbCr channels.
Step 2: Decompose the Y channel into ten DWT subbands.
Step 3: Re-fetch the stored authenticated key.
Step 4: Quantize absolute values of LL3 subband.
Step 5: Extract four least significant bits of re-fetched key.

## 5. EXPERIMENTAL RESULTS

The proposed perceptual watermarking framework was implemented for evaluating both properties of imperceptibility and robustness in a high payload.
Three 256×256 images: Lena, Peppers and Arm, shown in Fig 3(a-c) were taken as the host images to embed a 30×30 binary watermark image, shown in Fig 3(d). For the entire test results in this paper, MATLAB R2007a software is used. Also for computing the wavelet transforms, 9-7 biorthogonal spline (Bspline) wavelet filter are used. Cause of use of B-spline function wavelet is that, B-spline functions, do not have compact support, but are orthogonal and have better smoothness properties than other wavelet functions [15].
After watermark embedding, we calculate the percentage of the payload as follows:

$$\% \ Payload = \frac{\sum_i \sum_j W'_{i,j}}{\sum_i \sum_j W_{i,j}} \times 100 \quad (3)$$

In our scheme, we can obtain a payload, equal to %20.
We also measure the similarity of original host image and watermarked images by the standard correlation coefficient (Corr) as [9]:

$$Correlatio = \frac{\sum (x - x')(y - y')}{\sqrt{(x - x')^2}\sqrt{(y - x')^2}} \quad (4)$$

Moreover, the peak signal-to-noise ratio (PSNR) was used to evaluate the quality of the watermarked image. The PSNR is defined as [9, 16]:

$$PSNR = 10 \log_{10} \frac{255^2}{MSE} \ (dB) \quad (5)$$

Where mean-square error (MSE) is defined as:

$$MSE = \frac{1}{mn} \sum_{i=1}^{m} \sum_{j=1}^{n} (h_{i,j} - h'_{i,j})^2 \quad (6)$$

Where $h_{i,j}$ and $h'_{i,j}$ are the gray levels of pixels in the host and watermarked images, respectively. The larger PSNR is, the better the image quality is. In general, a watermarked image is acceptable by human perception if its PSNR is greater than 30 dBs. In other words, the correlation is used for evaluating the robustness of watermarking technique and the PSNR is used for evaluating the transparency of watermarking technique [9].
We also used the normalized correlation (NC) coefficient to measure the similarity between original watermarks W and the extracted watermarks W' that is defined as [17]:

$$NC = \frac{\sum_i \sum_j w_{i,j} * w'_{i,j}}{\sum_i \sum_j w_{i,j}^2} \quad (7)$$

The proposed watermarking approach yields satisfactory results in watermark imperceptibility and robustness. The PSNRs of the watermarked images produced by the proposed approach are all greater than 37.522 dBs, NCs between original watermark images and extracted watermark images are all equal 1, and correlations between host images and watermarked images are all greater than 0.999, which are perceptually imperceptible as shown in Fig 4.

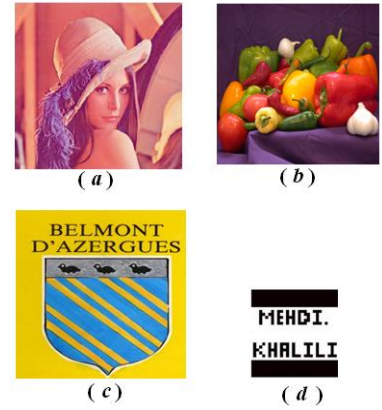

**Fig3. The host images for watermarking.
(a-c) Lena, Peppers, and Arm. (d) Watermark image**

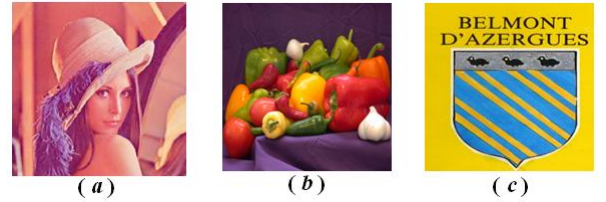

**Fig4. The watermarked images by the proposed approach**

Table 1 shows the extracting results and the watermarked images using the proposed method without any attacks. As it is visible the watermarked results are excellent.

**Table 1. Obtained results of watermark extracting**

| Image | Extracted Watermark | PSNR (dB) | Corr | NC | Error Bits |
|---|---|---|---|---|---|
| Lena | MEHDI. KHALILI | 37.5924 | 0.9996 | 1.00 | 0 |
| Peppers | MEHDI. KHALILI | 39.9783 | 0.9995 | 1.00 | 0 |
| Arm | MEHDI. KHALILI | 39.9046 | 0.9996 | 1.00 | 0 |

## 5.1. Robustness Examinations

After able to achieve the desired fidelity, watermarks robustness and their watermarked images detectability evaluate under compression and cropping attacks.

*5.1.1 Robustness to Compression Attacks*
We here examine the watermark robustness with wavelet compression. We compress the watermarked images with different compression thresholds using wavelet compression.

Then, we extract the watermark from the compressed watermarked images. The compression results are illustrated in Table 2. From the table, we can find that proposed method yields satisfactory imperceptible watermarked images. All PSNRs are greater than 37 dB, and all NC are greater than 0.83 under different compression thresholds of equal or less than 6.

**Table 2. Obtained results of wavelet compression attacks with different compression thresholds**

Threshold = 1.0

| Image | Lena | Peppers | Arm |
|---|---|---|---|
| Extracted Watermark | MEHDI. KHALILI | MEHDI. KHALILI | MEHDI. KHALILI |
| PSNR (dB) | 37.5923 | 39.9781 | 39.9045 |
| NC | 1.00 | 1.00 | 1.00 |
| Error Bits | 0 | 0 | 0 |

Threshold = 3.0

| Image | Lena | Peppers | Arm |
|---|---|---|---|
| Extracted Watermark | MEHDI. KHALILI | MEHDI. KHALILI | MEHDI. KHALILI |
| PSNR (dB) | 37.5918 | 39.9778 | 39.9042 |
| NC | 0.9777 | 0.9891 | 0.9933 |
| Error Bits | 10 | 8 | 3 |

Threshold = 6.0

| Image | Lena | Peppers | Arm |
|---|---|---|---|
| Extracted Watermark | MEHDI. KHALILI | MEHDI. KHALILI | MEHDI. KHALILI |
| PSNR (dB) | 37.5905 | 39.9762 | 39.9033 |
| NC | 0.8392 | 0.8550 | 0.8951 |
| Error Bits | 72 | 65 | 47 |

*5.1.2 Robustness to Cropping Attacks*

In this examination, we crop two different areas of the watermarked images and then, from the cropped images, we extract the watermark image. The results of the extracted watermarks are shown in Table 3. In this table the cropped areas of watermarked imaged and their extracted watermarks are shown, too. As it is obvious, NCs are still greater than 0.87.

## 6. WATERMARKING PROCESS IN RGB COLOR SPACE

To consideration of proposed technique efficiency, we here perform watermarking process in RGB color space by the same method. Then we compare the obtained results of watermark extracting and robustness examinations shown in tables 4, 5, 6, with the previous obtained results.

From the table 4, it can shown that transparency of watermarking technique in YCbCr color space is better than in RGB color space, because of its PSNRs.

Also tables 5 and 6 show that, robustness to compression attacks, in YCbCr color space is obviously greater than in RGB color space. In spite of these advantages, robustness to cropping attacks, in YCbCr color space is less than in RGB color space.

Furthermore, computing complication in YCbCr color space is greater than in RGB color space, too.

**Table 3. Obtained results of cropping attacks**

Cropped Area 1

| Image | Lena | Peppers | Arm |
|---|---|---|---|
| Cropped Image | | | |
| Extracted Watermark | MEH | MEH | MEH |
| NC | 0.8723 | 0.9454 | 0.9239 |
| Error Bits | 355 | 362 | 356 |

Cropped Area 2

| Image | Lena | Peppers | Arm |
|---|---|---|---|
| Cropped Image | | | |
| Extracted Watermark | LIL | LIL | LIL |
| NC | 0.8725 | 0.9541 | 0.7995 |
| Error Bits | 357 | 353 | 364 |

**Table 4. Obtained results of watermark extracting in RGB color space**

| Image | Extracted Watermark | PSNR (dB) | Corr | NC | Error Bits |
|---|---|---|---|---|---|
| Lena | MEHDI. KHALILI | 34.1554 | 0.9989 | 1.00 | 0 |
| Peppers | MEHDI. KHALILI | 35.5328 | 0.9991 | 1.00 | 0 |
| Arm | MEHDI. KHALILI | 33.7440 | 0.9987 | 1.00 | 0 |

## 7. CONCLUSION

We have proposed a watermarking framework for embedding visually recognizable binary watermark in color images, which can resist image-processing attacks, especially the wavelet compression attacks. In most DWT-based watermarking frameworks, watermarks are often in the form of random-number sequences or gray-level images. In this paper, we proposed an image accreditation technique by embedding binary image watermark into color images.

In the proposed approach, a host image was converted into YCbCr color space using the CCIR 601 standard, and then, the Y channel was decomposed into wavelet coefficients.

Then, the watermark was embedded into the four least significant bits of host coefficients in the approximation coefficients subband, using LSB insertion technique. The experimental results show that the proposed method provides extra imperceptibility and robustness of watermarking against wavelet compression attacks compared to the traditional methods in RGB color space. Moreover, the proposed approach has no need of the original host image to extract watermarks.

**Table 5. Obtained results of wavelet compression attacks with different compression thresholds in RGB color space**

Threshold = 1.0

| Image | Lena | Peppers | Arm |
|---|---|---|---|
| Extracted Watermark | 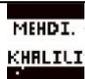 | 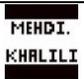 | 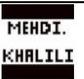 |
| PSNR (dB) | 34.1215 | 35.4946 | 33.7186 |
| NC | 0.9933 | 1.00 | 1.00 |
| Error Bits | 3 | 0 | 0 |

Threshold = 3.0

| Image | Lena | Peppers | Arm |
|---|---|---|---|
| Extracted Watermark | 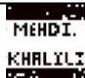 | 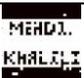 | 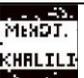 |
| PSNR (dB) | 32.6441 | 34.5888 | 32.6405 |
| NC | 0.9251 | 0.9477 | 0.9280 |
| Error Bits | 35 | 24 | 33 |

Threshold = 6.0

| Image | Lena | Peppers | Arm |
|---|---|---|---|
| Extracted Watermark | 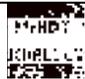 | 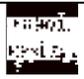 | 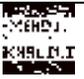 |
| PSNR (dB) | 30.9761 | 30.1547 | 30.5760 |
| NC | 0.7376 | 0.8346 | 0.8026 |
| Error Bits | 132 | 79 | 94 |

**Table 6. Obtained results of cropping attacks in RGB color space**

Cropped Area 1

| Image | Lena | Peppers | Arm |
|---|---|---|---|
| Cropped Image | 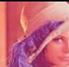 | 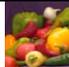 | 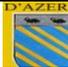 |
| Extracted Watermark | 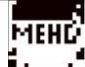 | 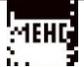 | 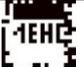 |
| NC | 0.9299 | 0.9615 | 0.9379 |
| Error Bits | 267 | 279 | 304 |

Cropped Area 2

| Image | Lena | Peppers | Arm |
|---|---|---|---|
| Cropped Image | 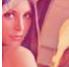 | 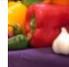 | 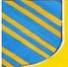 |
| Extracted Watermark | 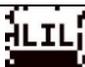 | 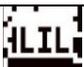 | 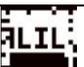 |
| NC | 0.9286 | 0.9664 | 0.8267 |
| Error Bits | 321 | 307 | 315 |